\documentclass[preprint,prd,aps,article,nofootinbib]{revtex4}
\usepackage{slashed}
\usepackage{tikz}
\usepackage{graphicx}
\usepackage{diagbox}
\usepackage{float}
\usepackage[colorlinks,
            linkcolor=red,
            anchorcolor=blue,
            citecolor=black
            ]{hyperref}
\usepackage{doi}
\usepackage{color}
\usepackage{slashed}
\begin{document}
\title{Production of $D_{s0}(2590)^+$ in $B$ nonleptonic decays}
\author{Shuang-Tao Wang$^{1,2,3}$\footnote{20218015005@stumail.hbu.edu.cn}, Su-Yan Pei$^{1,2,3}$, Tianhong Wang$^{4}$, Guo-Li Wang$^{1,2,3}$\footnote{wgl@hbu.edu.cn, Corresponding author}}

\affiliation{$^1$ Department of Physics, Hebei University, Baoding 071002, China
\nonumber\\
$^{2}$ Hebei Key Laboratory of High-precision Computation and Application of Quantum Field Theory, Baoding 071002, China
\nonumber\\
$^{3}$ Hebei Research Center of the Basic Discipline for Computational Physics, Baoding 071002, China
\nonumber\\
$^{4}$ School of Physics, Harbin Institute of Technology, Harbin 150001, China}
\begin{abstract}
In 2021, a new charm-strange meson, $D_{s0}(2590)^+$, has been discovered, it is believed to be the $D_s^+(2^1S_0)$. However, its low mass and wide width are challenged by theoretical results. Given the small branching ratio of the current production channel, resulting in a small number of events and large errors. We suggest to search for the $D_{s0}(2590)^+$ in the $B$ meson nonleptonic decays, $B_q\rightarrow D^{(*)}_qD_{s0}(2590)^+$ ($q=u,d$), followed by $D_{s0}(2590)^+\to D^*K$. We find $Br(B_q\rightarrow D^{(*)}_qD_{s0}(2590)^+)\times Br(D_{s0}(2590)^+\to D^{*}K)=(2.16\sim2.82)\times 10^{-3}$ is very large, and the result is not sensitive to the mass of $D_{s0}(2590)^+$. Due to large branching ratio, large amount of $D_{s0}(2590)^+$ events are expected. This study is based on the framework of instantaneous Bethe-Salpeter equation, and the used relativistic wave functions for mesons contain different partial waves. The contributions of different partial waves are also studied.
\end{abstract}
\maketitle
\section{Introduction}
In the year of 2021, the LHCb collaboration discovered a new charmed-strange meson $D_{s0}(2590)^+$ in the decay process of $B^0\rightarrow D^-D^+K^+\pi^-$  \cite{[1]}. Its mass, width, and spin-parity are measured, and their values are $m=2591\pm6\pm7$ MeV, $\Gamma=89\pm16\pm12$ MeV and $J^P=0^-$, respectively. This new particle is believed to be the $D_s^+(2^1S_0)$ state, the first radial excitation of the pseudoscalar ground-state $D_s^+$ meson.

Once discovered, the $D_{s0}(2590)^+$ has attracted a lot of attention because as the radial excited state $D_s^+(2^1S_0)$, its mass is much lower than all the theoretical predictions, at least several tens of MeV \cite{[6],[7],[8],GM,Ebert,lidm}. Another issue is its width, on the surface, some theoretical results are consistent with experimental data. However, we pointed out in Ref.\cite{[2]} that due to its main Okubo-Zweig-Iizuka(OZI)-allowed two body strong decays occurring near the threshold, and because $\Gamma\propto {|\mathbf{P}_f}|^3$ ($\mathbf{P}_f$ is the three-dimensional recoil momentum), its width will strongly depend on its mass. Therefore, with different $D_{s0}(2590)^+$ masses, the calculated widths should not be directly compared to each other. To address this issue, we introduced an almost mass independent quantity. And our calculations showed that, unlike surface phenomena, all the theoretical predicted widths are much smaller than experimental data \cite{close,liux,GM,[14]}.

Ref.\cite{[16]} confirms our results, using a semi-relativistic model, they find the mass of $D_s^+(2^1S_0)$ is about 60 MeV larger than data, and predict the decay width to be $\Gamma(D_{s0}(2590)^+)\simeq19$ MeV, consistent with our result of $19.7$ MeV \cite{[2]}.
Applying the $3P^0$ model, Ref.\cite{[18]} estimates the width of $D_s^+(2^1S_0)$ is also about 20 MeV. To understand the $D_{s0}(2590)^+$, they take into account the coupling of $D^*K$, which can lowered the mass by about 88 MeV \cite{[18]}. In a coupled-channel framework, Ref.\cite{Ortega} confirms that to meet the assignment of the $D_{s0}(2590)^+$ as the $D_s^+(2^1S_0)$, the nearby meson-meson thresholds must be taken into account. From the current status, it can be seen that $D_{s0}(2590)^+$ is not well understood, so more careful theoretical and experimental studies on it are still needed. More comments on $D_{s0}(2590)^+$ can be found in the review of Ref.\cite{zhusl}.

Currently, the discovered channel of $D_{s0}(2590)^+$ is $B^0\to D^-D^+K^+\pi^-$, and $D_{s0}(2590)^+$ is reconstructed in the $D^+K^+\pi^-$ final state \cite{[1]}, which is a three-body strong decay of $D_{s0}(2590)^+$, and has a very small branching ratio resulting in a very limited $D_{s0}(2590)^+$ events and with large uncertainties  \cite{pdg}. While the OZI-allowed two-body strong decay $D_{s0}(2590)^+\to D^*K$ has a dominant branching ratio, therefore, instead of $B^0\to D^-D^+K^+\pi^-$, large amount of events and small errors are expected in $B^0\to D^-D^*K$.
So in this paper, we will further study the properties of $D_{s0}(2590)^+$ as the $D_s^+(2^1S_0)$ meson, and calculate its production rate in $B$ meson weak decays of $B_q\rightarrow D_q^{(*)} D_{s0}(2590)^+$ with $q=u,d$, followed by $D_{s0}(2590)^+\to D^*K$.


The non-leptonic decays $B_q\rightarrow D^{(*)}_q D_{s0}(2590)^+$ are calculated using the factorization assumption. And we consider not only the contribution of the leading color-allowed tree diagram but also the some others like the penguin diagram.
When calculating the decay amplitude, the instantaneous Bethe-Salpeter equation \cite{[25]} method, also called the Salpeter equation \cite{[24]} method, is chosen.
The advantage of this approach is that it is a relativistic method \cite{[3],[4]}, and the relativistic wave function contains rich information, for example, in addition to the common nonrelativistic main wave, there are also other relativistic partial waves. Besides the decay branching ratio, we will also study the behavior of different partial waves in the hadronic transition.

This paper is organized as follows, in Sec. II, taking $B^+\rightarrow \bar{D}^0D_{s0}(2590)^+$ as an example, we show how to calculate the transition matrix element, and give the wave functions with different partial waves. In Sec. III, we give our results and some discussions.
\section{Theoretical method}
\subsection{Decay amplitude}
According to the effective Hamiltonian \cite{[5]},  the amplitude of the non-leptonic decay $B^+\rightarrow \bar{D}^0D_{s0}(2590)^+$ can be written as \cite{[12]},
\begin{eqnarray}\label{coeff}
\mathcal{M}=\frac{G_F}{\sqrt{2}}\{{V_{cb}V_{cs}^*}{a_1}+\sum_{q'=u,c}V_{q'b}V_{q's}^*[a_{4}^{q'}+a_{10}^{q'}+\xi(a_{6}^{q'}+a_{8}^{q'})]\}A,
\end{eqnarray}
where $G_F$ is the Fermi constant; $V_{cb}$, $V_{cs}$, $V_{ub}$ and $V_{us}$ are the Cabibbo-Kobayashi-Maskawa (CKM) matrix elements \cite{pdg};
$a_i$ ($i=1$,4,6,8,10) are the combinations of Willson coefficients, their formulas and the expression of coefficient $\xi$ are shown in the Appendix; $A$ is the hadronic matrix element, in  the factorization assumption, it is written as
\begin{eqnarray}
A=<\bar{D}^0{\mid}J_{\mu}{\mid}B^+><D_{s0}(2590)^+{\mid}J^{\mu}{\mid}0>,
\end{eqnarray}
where $<D_{s0}(2590)^+{\mid}J^{\mu}{\mid}0>=iF_PP_{f_2}^{\mu}$, $F_P$ and $P_{f_2}$ are the decay constant and momentum of $D_{s0}(2590)^+$, respectively.

In the instantaneous approximation, the hadronic transition matrix element can be written as an overlapping integral with respect to the meson wave functions \cite{[4]},
\begin{eqnarray}\label{o}
<\bar{D}^0(P_f){\mid}J_{\mu}{\mid}B^+(P)>=\int\frac{d^{3}\vec{q}}{(2\pi)^{3}}Tr[\overline{\varphi}^{++}_{P_{f}}(\vec{q}_f)\frac{\slashed{P}}{M}\varphi^{++}_P(\vec{q})\gamma_{\mu}(1-\gamma_5)],
\end{eqnarray}
where $\varphi_{P}^{++}$ and $\overline{\varphi}_{P_f}^{++}=\gamma_0(\varphi_{P_f}^{++})^\dag\gamma^0$ are the positive-energy wave functions of the $B^+$ and $\bar{D}^0$ mesons, respectively; 
$P$ and $P_f$ are the momenta of $B^+$ and $\bar{D}^0$, respectively; $M$ is the mass of $B^+$; 
$\vec{q}$ and $\vec{q}_f$ are the relative momenta between quarks in $B^+$ and $\bar{D}^0$, respectively, and their relation is $\vec{q}_f=\vec{q}-\alpha'_1{\vec{P}_f}$, 
where $\alpha'_1=m_u/(m_u+m_c)$, $m_u$ and $m_c$ are the quark masses.

The formula for the decay width of the two-body is:
\begin{eqnarray}
\Gamma=\frac{\mid{\vec{P_f}}\mid}{8\pi{M^2}}\sum|\mathcal{M}|^2.
\end{eqnarray}
\subsection{Relativistic wave functions and their partial waves}
In our method, the relativistic wave function of a meson is provided according to its $J^P$, and the numerical values of radial parts are obtained by solving the Salpeter equation. The wave function obtained this way contains different partial waves.
\subsubsection{$0^-$ state}
In the instantaneous approximation $(P\cdot{q}=0)$, the relativistic wave function of a $0^-$ meson is written as \cite{[9]},
\begin{eqnarray}\label{0-}
\varphi_{0^-}(q_\perp)=(\slashed{P}f_1+Mf_2+\slashed{q}_{\perp}f_3+\frac{\slashed{P}\slashed{q}_{\perp}}{M}f_4)\gamma_5,
\end{eqnarray}
where $q_\perp=q-\frac{P\cdot q}{M^2}P=(0,\vec{q})$; the radial wave functions $f_i(i=1,2,3,4)$ are functions of $\vec{q}^2$. The corresponding positive-energy wave function is
\begin{eqnarray}\label{0-2}
\varphi^{++}_{0^{-}}=[\slashed{P}A_{1}+MA_{2}+\slashed{q}_{\perp}A_{3}+\frac{\slashed{P}\slashed{q}_{\perp}}{M}A_{4}]\gamma_{5},
\end{eqnarray}
where $A_{i}(i=1,2,3,4)$ are related to the radial wave functions $f_i$, and are shown in the Appendix. It can be verified that each term in Eq.(\ref{0-}) or Eq.(\ref{0-2}) has a quantum number of $0^-$ \cite{[10]}. We note that $(\slashed{P}A_{1}+MA_{2})\gamma_{5}$ are $S$-wave, and the $A_3$, $A_4$ terms are $P$-wave \cite{[10]}. In nonrelativistic limit, only $S$-wave exist, so $A_3$ and $A_4$ terms are relativistic corrections.
\subsubsection{$1^-$ state}
The relativistic wave function of the vector $1^-$ state is written as \cite{[11]},
\begin{eqnarray}
\varphi_{1^-}(q_{f_\perp})\nonumber&=&(q_{f_\perp}\cdot{\epsilon_f})\left[g_1+\frac{\slashed{P}_{f}}{M_{f}}g_2+\frac{\slashed{q}_{f_{\perp}}}{M_{f}}g_3+\frac{\slashed{P}_{f}\slashed{q}_{f_{\perp}}}{M_{f}^{2}}g_4
\right]+M_f\slashed{\epsilon}_f\left[g_5+\frac{\slashed{P}_{f}}{M_{f}}g_6\right]\\&+&(\slashed{q}_{f_\perp}\slashed{\epsilon}_f-q_{f_\perp}\cdot{\epsilon_f})g_7+\frac{1}{M_f}
(\slashed{P}_{f}\slashed{\epsilon}_f\slashed{q}_{f_{\perp}}-\slashed{P}_{f}q_{f_\perp}\cdot\epsilon_f)g_8,
\end{eqnarray}
where $\epsilon_f$ is the polarization vector; $g_i(i=1,2,...)$ are functions of ${\vec q}_{f}^2$. The positive-energy wave function is written as,
\begin{eqnarray}\label{1-}
\varphi^{++}_{1^{-}}\nonumber&=&q_{f_{\perp}}\cdot\epsilon_{f}\left[C_{3}+C_{4}\frac{\slashed{P}_{f}}{M_{f}}+C_{7}\frac{\slashed{q}_{f_{\perp}}}{M_{f}}+C_{8}
\frac{\slashed{P}_{f}\slashed{q}_{f_{\perp}}}{M_{f}^{2}}\right]\\
&+&M_{f}\slashed{\epsilon}_{f}\left[C_{1}+C_{2}\frac{\slashed{P}_{f}}{M_{f}}+C_{5}\frac{\slashed{q}_{f_{\perp}}}{M_{f}}+C_{6}\frac{\slashed{P}_{f}\slashed{q}_{f_{\perp}}}{M_{f}^{2}}\right],
\end{eqnarray}
where $C_i(i=1,2,3,\ldots)$ are functions of $g_i$, and their expressions are shown in the Appendix.

In Eq.(\ref{1-}), $C_1$ and $C_2$ terms are nonrelativistic $S$-wave, all others are relativistic corrections;  $C_3$, $C_4$, $C_5$ and $C_6$ are $P$-wave; $C_7$ and $C_8$ terms are $D$-wave mixed with $S$-wave, the pure $D$-wave is \cite{[15]}
\begin{eqnarray}
\left[(q_{f_{\perp}}\cdot\epsilon_{f})
\slashed{q}_{f_\perp}-\frac{1}{3}q^2_{f_{\perp}}\slashed{\epsilon}_{f}\right]\left[C_7\frac{1}{M_f}-C_{8}
\frac{\slashed{P}_{f}}{M_{f}^{2}}\right],
\end{eqnarray}
and the corresponding $S$-wave is
\begin{eqnarray}\label{S'}
\frac{1}{3}q^2_{f_{\perp}}\slashed{\epsilon}_{f}
\left[C_7\frac{1}{M_f}-C_{8}
\frac{\slashed{P}_{f}}{M_{f}^{2}}\right].
\end{eqnarray}
\subsection{Form factors}
Using the positive-energy wave functions, the transition matrix element in Eq.(\ref{o}) is calculated straightly. And the transition matrix elements for the transitions $0^{-}\rightarrow0^{-}$ and $0^{-}\rightarrow 1^{-}$ can be expressed as functions of form factors,
\begin{eqnarray}
<\bar{D}^0{\mid}J_{\mu}{\mid}B^+>&=&t_1P_{\mu}+t_2P_{f_{\mu}},\\
<\bar{D}^{*0}{\mid}J_{\mu}{\mid}B^+>&=&id_1\varepsilon_{{\mu}\epsilon_{f}PP_{f}}+\epsilon_f\cdot{P}(d_2P_{\mu}+d_3P_{f_{\mu}})+d_4\epsilon_{f_{\mu}},
\end{eqnarray}
where $t_1$, $t_2$, $d_{1}$, $d_{2}$, $d_{3}$ and $d_{4}$ are the form factors. In the upper formula, we have used the following shorthand notation $\varepsilon_{{\mu}{\nu}\alpha\beta}\epsilon_f^\nu P^\alpha P_f^\beta=\varepsilon_{{\mu}\epsilon_{f}PP_{f}}$.
\section{Numerical results and discussion}
The following quark masses, $m_b=4.96$ GeV, $m_c=1.62$ GeV, $m_u=0.305$ GeV, $m_d=0.311$ GeV and $m_s=0.5$ GeV are chosen. For the decay constant of $D_s^+(1968)$, we take $f_{D_s^+(1968)}=0.25$ GeV from PDG \cite{pdg}, while the decay constant of $D_{s0}(2590)^+$, $f_{D_{s0}(2590)^+}=0.20$ GeV is calculated by our model.

\subsection{$B_q\rightarrow D^{(*)}_qD_s^+(1968)$}
To validate our method, we first study the processes $B_q\rightarrow D^{(*)}_qD_s^+(1968)$ and show the decay branching ratios in Table~\ref{2}.
\begin{table}[!h]
\begin{center}
\caption{The branching ratios $(10^{-3})$ of $B_q\rightarrow D_q^{(*)}D_s^+(1968)$}
\begin{tabular}{ccc}
\hline
\hline
Decay~channel&Ours&~~~~PDG\cite{pdg}\\
\hline
$B^+\rightarrow\bar{D}^{0}D^+_s$&9.60&~~$10.0^{+1.7}_{-1.7}$\\
$B^+\rightarrow\bar{D}^{*0}D^+_s$&10.7&~~$8.2^{+1.7}_{-1.7}$\\
\hline
$B^0\rightarrow D^-D^+_s$&8.90&~~$7.2^{+0.8}_{-0.8}$\\
$B^0\rightarrow D^{*-}D^+_s$&9.91&~~$8.0^{+1.1}_{-1.1}$\\
\hline
\hline
\end{tabular}\label{2}
\end{center}
\end{table}
Our results are very close to experimental data, which shows that our method is feasible. Therefore, we apply it to $B_q\rightarrow D^{(*)}_qD_{s0}(2590)^+$.
\subsection{$B_q\rightarrow D^{(*)}_qD_{s0}(2590)^+$}
The calculated branching ratios of decays $B_q\rightarrow D_q^{(*)}D_{s0}(2590)^+$ are shown in Table~\ref{1}.
\begin{table}[!h]
\begin{center}
\caption{The branching ratios $(10^{-3})$ of $B_q\rightarrow D_q^{(*)}D_{s0}(2590)^+$}
\begin{tabular}{ccc}
\hline
\hline
Decay~channel&~~~~~~~~Br~~~~~~~\\
\hline
$B^+\rightarrow\bar{D}^{0}D_{s0}(2590)^+$&~4.96\\
$B^+\rightarrow\bar{D}^{*0}D_{s0}(2590)^+$&~5.35\\
\hline
$B^0\rightarrow D^-D_{s0}(2590)^+$&~4.58\\
$B^0\rightarrow D^{*-}D_{s0}(2590)^+$&~4.90\\
\hline
\hline
\end{tabular}\label{1}
\end{center}
\end{table}
Compared with the results in Table~\ref{2}, we can see that the branching ratios of $B_q\rightarrow D_q^{(*)}D_{s0}(2590)^+$ are very large, about half of that of $B_q\rightarrow D_q^{(*)}D_s^+(1968)$. The differences mainly come from the differences in phase spaces and decay constants.

In Ref.\cite{[2]}, we obtained $$\Gamma(D_{s0}(2590)^+\to D^*(2007)^0K^+)=10.4 ~\rm{MeV},$$ $$\Gamma(D_{s0}(2590)^+\to D^*(2010)^+K^0)=9.29 ~\rm{MeV}.$$ Ignoring other secondary contributions, the total width can be estimated as the sum of them,
$\Gamma(D_{s0}(2590)^+) \simeq 19.7 ~\rm{MeV}$. Then we obtain the following results
\begin{eqnarray}\label{ratio1}
Br(B^+\rightarrow\bar{D}^{0}D_{s0}(2590)^+)\times Br(D_{s0}(2590)^+\to D^{*0}K^+)=2.62\times 10^{-3},
\nonumber \\
Br(B^+\rightarrow\bar{D}^{0}D_{s0}(2590)^+)\times Br(D_{s0}(2590)^+\to D^{*+}K^0)=2.34\times 10^{-3},
\nonumber \\
Br(B^+\rightarrow\bar{D}^{*0}D_{s0}(2590)^+)\times Br(D_{s0}(2590)^+\to D^{*0}K^+)=2.82\times 10^{-3},
\nonumber \\
Br(B^+\rightarrow\bar{D}^{*0}D_{s0}(2590)^+)\times Br(D_{s0}(2590)^+\to D^{*+}K^0)=2.53\times 10^{-3},
\end{eqnarray}
\begin{eqnarray}\label{ratio2}
Br(B^0\rightarrow{D}^{-}D_{s0}(2590)^+)\times Br(D_{s0}(2590)^+\to D^{*0}K^+)=2.42\times 10^{-3},
\nonumber \\
Br(B^0\rightarrow{D}^{-}D_{s0}(2590)^+)\times Br(D_{s0}(2590)^+\to D^{*+}K^0)=2.16\times 10^{-3},
\nonumber \\
Br(B^0\rightarrow{D}^{*-}D_{s0}(2590)^+)\times Br(D_{s0}(2590)^+\to D^{*0}K^+)=2.59\times 10^{-3},
\nonumber\\
Br(B^0\rightarrow{D}^{*-}D_{s0}(2590)^+)\times Br(D_{s0}(2590)^+\to D^{*+}K^0)=2.31\times 10^{-3}.
\end{eqnarray}

Since the mass of $D_{s0}(2590)^+$ is lower than all the theoretical predictions, we adjust its mass in the range of $2600\sim2670$ MeV, and show the corresponding branching ratio of $B_q\rightarrow D_q^{(*)}D_{s0}(2590)^+$ in Table~\ref{3}. It can be seen that as the $D_{s0}(2590)^+$ mass increases, the branching ratio decreases. But due to the large phase space, the decay branching ratio is not very sensitive to the mass of $D_{s0}(2590)^+$.

\begin{table}[!h]
\begin{center}
\caption{The branching ratios $(10^{-3})$ vary with the mass of $D_{s0}(2590)^+$}
\begin{tabular}{ccccccccc}
\hline
\hline
$M_{D_{s0}(2590)^+}$&~~$2600$~~&~~$2610$~~&~~$2620$~~&~~$2630$~~&~~$2640$~~&~~$2650$~~&~~$2660$~~&~~$2670$~~\\
\hline
$Br(B^+\rightarrow\bar{D}^{0}D_{s0}(2590)^+)$&4.95&4.92&4.89&4.86&4.83&4.80&4.76&4.73\\
$Br(B^+\rightarrow\bar{D}^{*0}D_{s0}(2590)^+)$&5.34&5.30&5.26&5.23&5.18&5.14&5.09&5.05\\
\hline
$Br(B^0\rightarrow D^-D_{s0}(2590)^+)$&4.58&4.55&4.52&4.50&4.47&4.44&4.40&4.37\\
$Br(B^0\rightarrow D^{*-}D_{s0}(2590)^+)$&4.89&4.85&4.82&4.79&4.75&4.71&4.67&4.63\\
\hline
\hline
\end{tabular}\label{3}
\end{center}
\end{table}

For the strong decays $D_{s0}(2590)^+\to D^{*0}K^+$ and $D_{s0}(2590)^+\to D^{*+}K^0$, they occur near the threshold of $D_{s0}(2590)^+$, so the decay widths strongly depend on the mass of $D_{s0}(2590)^+$ \cite{[2]}. However, the corresponding branching ratios are not sensitive to the mass \cite{[2]}, so we conclude that the results in Eqs.(\ref{ratio1}) and (\ref{ratio2}) are not sensitive to the mass of $D_{s0}(2590)^+$.

\subsection{Contributions of different partial waves in hadron transition}
The wave functions we provided contain different partial waves, so it is interesting to see their behaviors in the decays. We take the processes $B^+\rightarrow \bar{D}^{(*)0}D_{s0}(2590)^+$ as examples to provide the details. Since the $D_{s0}(2590)^+$ only contributes to the decay constant, we ignore the discussion of its partial waves and only consider the contribution of partial waves in the transition $B^+ \rightarrow \bar{D}^{(*)0}$.
\subsubsection{$B^+ \rightarrow \bar{D}^0D_{s0}(2590)^+$}
We show the results in Table~\ref{4}. Where the ``whole" means the contribution using the complete wave function, while the ``$S$-wave" means the result is obtained only using the $S$-wave and ignoring others, etc. And we use the ``prime" to represent the final state.
\begin{table}[!h]
\begin{center}
\caption{Contributions of partial waves to the branching ratio of $B^+$$\rightarrow$$\bar{D}^{0}$$D_{s0}(2590)^+$ ($10^{-3}$)}
\begin{tabular}{|c|c|c|c|}
\hline
{\diagbox{$B^+$}{$\bar{D}^{0}$}}&whole&~~$S'$-wave ($A'_1, A'_2$)~~&~~$P'$-wave ($A'_3, A'_4$)~~\\
\hline
whole&$4.96$&$5.12$&$0.0012$\\
\hline
$S$-wave ($A_1, A_2$)&$2.87$&$4.02$&$0.098$\\
\hline
$P$-wave ($A_3, A_4$)&$0.285$&$0.066$&$0.077$\\
\hline
\end{tabular}\label{4}
\end{center}
\end{table}
Table~\ref{4} shows that, the $S\times S'$ contribution to the branching ratio is the largest,  others are small. In the nonrelativistic limit, only $S\times S'$ exists, its contribution is $Br_{non}=4.02\times10^{-3}$. And all others provide the relativistic corrections. Our relativistic result is $Br_{rel}=4.96\times10^{-3}$, so the relativistic effect in transition $B^+ \rightarrow \bar{D}^0$ can be estimated as
\begin{eqnarray}\label{effect1}
\frac{Br_{rel}-Br_{non}}{Br_{rel}}=19.0\%.
\end{eqnarray}
\subsubsection{$B^+ \rightarrow \bar{D}^{*0}D_{s0}(2590)^+$}
\begin{table}[!h]
\begin{center}
\caption{Contributions of partial waves to the branching ratio of $B^+$$\rightarrow$$\bar{D}^{*0}$$D_{s0}(2590)^+$ ($10^{-3})$}
\begin{tabular}{|c|c|c|c|c|}
\hline
{\diagbox{$B^+$}{$\bar{D}^{*0}$}}&whole&~~$S'$-wave~~&~~$P'$-wave ($C_3, C_4,C_5,C_6$)~~&~~$D'$-wave~~\\
\hline
whole&$5.35$&$3.89$&$0.098$&$1.15\times10^{-3}$\\
\hline
$S$-wave ($A_1, A_2$)&$3.07$&$3.59$&$0.030$&$8.52\times10^{-4}$\\
\hline
$P$-wave ($A_3, A_4$)&$0.304$&$6.06\times10^{-3}$&$0.235$&$2.23\times10^{-5}$\\
\hline
\end{tabular}\label{5}
\end{center}
\end{table}
Table~\ref{5} shows the contribution of partial waves in $B^+$ and $\bar{D}^{*0}$ to the branching ratio of  $B^+$$\rightarrow$$\bar{D}^{*0}$$D_{s0}(2590)^+$.
The $S\times S'$ has the largest contribution, $3.59\times10^{-3}$, but this is not the nonrelativistic result. When calculating the nonrelativistic result, we should delete the contribution of the $S'$-wave from $C_7$ and $C_8$ terms in Eq.(\ref{S'}), and we obtain $Br_{non}=3.83\times10^{-3}$. Then the relativistic effect is
\begin{eqnarray}\label{effect2}
\frac{Br_{rel}-Br_{non}}{Br_{rel}}=28.4\%.
\end{eqnarray}
Comparing Eqs.(\ref{effect1}) and (\ref{effect2}), the relativistic effects are slightly different, but comparable. However, there are significant differences in the details of relativistic corrections in Tables~\ref{4} and \ref{5}.
In Table~\ref{4}, the relativistic contributions of $P\times P'$, $S\times P'$ and $P\times S'$ are comparable, while the corresponding ones in Table~\ref{5} are very different, with the following relationship: $P\times P' \gg S\times P'\gg P\times S'$. The contribution of $D'$-wave in $\bar{D}^{*0}$ is very small.
\section{Conclusion}
In the framework of the Salpeter equation, we study the production rates of $D_{s0}(2590)^+$ in the nonleptonic decays of $B$ mesons, and the branching ratios are $Br(B^+\rightarrow\bar{D}^{0}D_{s0}(2590)^+)=4.96\times10^{-3}$, $Br(B^+\rightarrow\bar{D}^{*0}D_{s0}(2590)^+)=5.35\times10^{-3}$,
$Br(B^0\rightarrow D^-D_{s0}(2590)^+)=4.58\times10^{-3}$,  $Br(B^0\rightarrow D^{*-}D_{s0}(2590)^+)=4.90\times10^{-3}$.
The $D_{s0}(2590)^+$ can be reconstructed in $D^{*}K$ invariant mass spectrum, for these cascading decays, we have $Br(D_{s0}(2590)^+\to D^{*0}K^+)=0.528$ and $Br(D_{s0}(2590)^+\to D^{*+}K^0)=0.472$. So given the large amount $B$-meson events in Belle and BaBar \cite{[19],[21]}, etc., it is expected that $D_{s0}(2590)^+$ will be carefully studied through these channels.

The quantum number $J^P$ of a meson determines that its wave function is not a pure wave, but contains different partial waves.
We studied the contributions of the partial waves in the transitions of $B^+\rightarrow\bar{D}^{0}$ and $B^+\rightarrow\bar{D}^{*0}$, and found that the main waves contribute to the nonrelativistic results, while other waves provide the relativistic corrections.

\textbf{Acknowledgments}

This work was supported in part by the National Natural Science Foundation of China (NSFC) under the Grants No. 12075073 and No. 12375085, the Natural Science Foundation of Hebei province under the Grant No. A2021201009.

\textbf{Appendix A: coefficients of the positive-energy wave function}

For the positive-energy wave function of  $0^-$ state, we have the following expression,
$A_{1}=\frac{1}{2}(f_{1}+f_{2}\frac{m_{1}+m_{2}}{\omega_{1}+\omega_{2}})$, $A_{2}=\frac{1}{2}(f_{2}+f_{1}\frac{\omega_{1}+\omega_{2}}{m_{1}+m_{2}})$,
$A_{3}=-A_{1}\frac{m_{1}-m_{2}}{m_{2}\omega_{1}+m_{1}\omega_{2}}$, $A_{4}=-A_{1}\frac{\omega_{1}+\omega_{2}}{m_{2}\omega_{1}+m_{1}\omega_{2}}$,\\
where $m_1$, $m_2$, $\omega_1=\sqrt{m_1^2+\vec{q}^2}$ and $\omega_2=\sqrt{m_2^2+\vec{q}^2}$ are the masses and energies of quark 1 and antiquark 2, respectively.

For the $1^-$ state, we have\\
$C_{1}=\frac{1}{2}\left[g_5-\frac{\omega_1+\omega_2}{m_1+m_2}g_6\right]$, $C_2=\frac{1}{2}\left[-\frac{m_1+m_2}{\omega_1+\omega_2}g_5+g_6\right]$,\\
$C_3=\frac{1}{2M_f(m_1\omega_2+m_2\omega_1)}[-(\omega_1+\omega_2)q^2_{f_\perp}g_3-(m_1+m_2)q^2_{f_\perp}g_4+2M_f^2\omega_2g_5-2M_f^2m_2g_6]$,\\
$C_4=\frac{1}{2M_f(m_1\omega_2+m_2\omega_1)}[-(m_1-m_2)q^2_{f_\perp}g_3-(\omega_1-\omega_2)q^2_{f_\perp}g_4-2M_f^2m_2g_5+2M_f^2\omega_2g_6]$,\\
$C_{5}=\frac{M_f}{2}\frac{\omega_1-\omega_2}{m_1\omega_2+m_2\omega_1}\left[g_5-\frac{\omega_1+\omega_2}{m_1+m_2}g_6\right]$, $C_{6}=\frac{M_f}{2}\frac{m_1+m_2}{m_1\omega_2+m_2\omega_1}\left[-g_5+\frac{\omega_1+\omega_2}{m_1+m_2}g_6\right]$,\\
$C_7=\frac{1}{2}\left[g_3+\frac{m_1+m_2}{\omega_1+\omega_2}g_4-\frac{2M_f^2}{m_1\omega_2+m_2\omega_1}g_6\right]$, $C_8=\frac{1}{2}\left[g_4+\frac{\omega_1+\omega_2}{m_1+m_2}g_3-\frac{2M_f^2}{m_1\omega_2+m_2\omega_1}g_5\right]$.

\textbf{Appendix B: some related formulas}

In Eq.(\ref{coeff}), for decay $B^+\rightarrow \bar{D}^0D_s^+$, we have
$\xi={2M_{D_s^+}^2}/(m_b-m_c)/(m_c+m_s)$, while for  $B^+\rightarrow \bar{D}^{*0}D_s^+$,  $\xi=-{2M_{D_s^+}^2}/(m_b+m_c)/(m_c+m_s)$. We also have
$a_{1}=c_{1}+{c_{2}}/{N_c}$, $a_{2i}=c_{2i}+{c_{2i}}/{N_c}$,
where $i=2,3,4,5$ and $N_c=3$, and the values of Willson coefficients $c_i$ are same as in the Ref. {\cite{[17]}}. The expression for $a_i^{q'}$ ($i=4,6,8,10$) is, $a_i^{q'}=a_i+I_i^{q'}$ with
$
I_4^{q'}=I_6^{q'}=\frac{\alpha_s}{9\pi}\left\{c_1\left[\frac{10}{9}-G(m_{q'},k^2)\right]\right\}
$,
$
I_8^{q'}=I_{10}^{q'}=\frac{\alpha_e}{9\pi}\frac{1}{N_c}\left\{(c_1+c_2N_c)\left[\frac{10}{9}-G(m_{q'},k^2)\right]\right\}
$,
where the coupling constants $\alpha_s=0.216$ and $\alpha_e={1}/{128}$;
$
G(m_{q'},k^2)=-4\int_0^1x(1-x)\ln\frac{m_{q'}^2-k^2x(1-x)}{m_b^2}dx
$, and the momentum $k$ is taken from Ref.\cite{[20]}.

\end{document}